# Atomic line defects and zero-energy end states in monolayer Fe(Te,Se) high-temperature superconductors

Cheng Chen[1], Kun Jiang[2], Yi Zhang[3], Chaofei Liu[1], Yi Liu[1], Ziqiang Wang[2] & Jian Wang[1,4,5,6]*

[1]*International Center for Quantum Materials, School of Physics, Peking University, Beijing 100871, China*

[2]*Department of Physics, Boston College, Chestnut Hill, Massachusetts 02467, USA*

[3]*Kavli Institute of Theoretical Sciences, University of Chinese Academy of Sciences, Beijing, 100190, China*

[4]*Collaborative Innovation Center of Quantum Matter, Beijing 100871, China*

[5]*CAS Center for Excellence in Topological Quantum Computation, University of Chinese Academy of Sciences, Beijing 100190, China*

[6]*Beijing Academy of Quantum Information Sciences, Beijing 100193, China*

*E-mail: jianwangphysics@pku.edu.cn*

**Majorana zero-energy bound states (ZEBSs) have been proposed to exist at the ends of one-dimensional Rashba nanowires proximity-coupled to an *s*-wave superconductor in an external magnetic field induced Zeeman field[1,2]. Such hybrid structures have been a central platform in the search for non-Abelian Majorana zero modes (MZMs) toward fault-tolerant topological quantum computing[3,4]. Here we report the discovery of ZEBSs simultaneously appearing at each end of a one-dimensional atomic line defect in monolayer iron-based high-temperature superconductor FeTe$_{0.5}$Se$_{0.5}$ films grown on SrTiO$_3$(001) substrates. The spectroscopic properties of the ZEBSs, including the temperature and tunneling barrier dependences, as well as their fusion induced by coupling on line defects of different lengths are found to be robust and consistent with those of the MZMs. These observations suggest a realization of topological Shockley defects at the ends of an atomic line defect in a two-dimensional *s*-wave superconductor that can host a Kramers pair of MZMs protected by time-reversal symmetry along the chain. Our findings reveal an unprecedented class of topological line defect excitations in two-dimensional superconductor FeTe$_{0.5}$Se$_{0.5}$ monolayer films and offer an advantageous platform for generating topological zero-energy excitations at higher operating temperatures, in a single material, and under zero external magnetic field.**

Zero-energy bound states (ZEBSs) localized at defects and ends of superconductors have attracted tremendous interests recently. Such exotic excitations, known as Majorana zero modes (MZMs) that are self-conjugate and obey non-Abelian statistics, have been shown theoretically to exist in the vortex core of certain p+ip topological superconductors[5,6] and at the ends of one-dimensional spinless p-wave topological superconductors[7]. MZMs can be used to construct nonlocal topological qubits which are robust against local perturbations as the basic building block for fault-tolerant topological quantum computing. While the search for intrinsic topological superconductors has been extremely challenging and yet unsuccessful, the recent interests in this direction have surged following the realization that producing the MZMs does not require an intrinsic topological superconductor. Fu and Kane proposed that when superconductivity is induced in the helical Dirac fermion surface states of a three-dimensional strong topological insulator by proximity coupling to an *s*-wave superconductor, a MZM would arise in the vortex core[8]. In the presence of Rashba spin-orbit coupling (RSOC), MZMs were predicted to arise at the two ends of a nanowire proximity coupled to an *s*-wave superconductor and a time-reversal symmetry breaking Zeeman field[1,2].

These proposals have fueled intensive experimental researches for MZMs in hybrid structures and materials without requiring topological superconductivity[9]. ZEBSs have been detected in the vortices of strong topological insulators and *s*-wave superconductor heterostructures by scanning tunneling microscopy[10]. Moreover, ZEBSs have been observed recently in a fraction of magnetic field induced vortices in the superconducting (SC) transition temperature ($T_c$) ~ 14.5 K bulk Fe(Te,Se) [11,12], hosting the observed SC Dirac cone topological surface states[13-16], as well as at the magnetic interstitial Fe impurities[17] where anomalous vortices may nucleate in zero external field[18]. The most advanced path currently is the nanowire/*s*-wave superconductor/Zeeman field hybrid structure[19-21] where a ZEBS has been detected at one end with a tunneling conductance close to the quantized value expected of a MZM[22-25]. Despite these promising results, the fabrication process, the low $T_c$, and the hard to control proximity effect present several basic challenges hindering the potential applications of the hybrid structures.

By using *in-situ* scanning tunneling microscopy/spectroscopy (STM/STS), we report here the discovery of robust ZEBSs at both ends of one-dimensional (1D) atomic line defects in two-dimensional (2D) one-unit-cell-thick $FeTe_{0.5}Se_{0.5}$ films grown on $SrTiO_3(001)$ substrates (1-UC $FeTe_{0.5}Se_{0.5}$/STO), which exhibit the $T_c$ around 62 K. These line defects naturally emerge during the growth process and correspond to lines of missing Te/Se atoms as shown in the schematic (Fig. 1e) (See additional discussions on the details of the line defects in the Methods). The significantly higher $T_c$ allows the ZEBSs to develop below temperatures around 24 K. At low temperatures, the tunneling spectra outside and near the middle section of a long defect chain are fully gapped, leaving only the zero-bias conductance peaks (ZBCPs) at the chain ends that decay rapidly in amplitude away from the ends without spatial dispersion. The ZBCP does not split with increasing tunneling barrier conductances and becomes sharper and higher as the tip approaches the film. Moreover, on shorter defect chains, we observe the coupling between the ZEBSs at their ends that leads to reduced ZBCPs even in the middle sections of the chain. These findings are consistent with the emergence of Majorana bound states at the ends of the 1D atomic line defects and reveal a superconducting analog of the Shockley surface/defect states in band theory[26]. Indeed, Shockley Majorana bound states have been proposed to exist at the ends of a 1D electrostatic line defect in a 2D chiral p-wave topological superconductor theoretically[27]. Although there is no evidence for 1-UC Fe(Te,Se) to be such a topological superconductor, we conjecture that the strong RSOC in even *s*-wave superconductors may produce a Kramers pair of MZMs protected by time-reversal symmetry at each end of the defect chain. Our discovery provides a promising and advantage single material platform to detect ZEBSs at the ends of 1D atomic line defects in high temperature superconductors, and opens a new direction for realizing Majorana bound states in solid state systems.

The high-quality 1-UC $FeTe_{0.5}Se_{0.5}$ film was epitaxially grown on a $SrTiO_3(001)$ substrate in ultrahigh vacuum (UHV) molecular beam epitaxy (MBE) system (see Methods for details). The nominal stoichiometry of $FeTe_{1-x}Se_x$ was calculated by measuring the thickness of second-layer $FeTe_{1-x}Se_x$ film and counting the ratio of Te/Se atoms in surface corrugation (Extended Data Fig. 1). In Fig. 1a, the STM topographic atomic-resolution image of 1-UC $FeTe_{0.5}Se_{0.5}$ film shows the topmost Te/Se atom arrangement with the typical characteristic of randomly distributed Se (darker) and Te (brighter) atoms. We perform the STS measurements of the tunneling conductance using the standard technique (see Methods for details of the calibration of zero-bias). The typical STS spectrum on the 1-UC $FeTe_{0.5}Se_{0.5}$ surface (Fig. 1b) exhibits the remarkable fully gapped superconductivity at 4.2 K. The two pairs of pronounced coherence peaks in the U-shaped spectrum indicate two superconducting gaps of $\Delta_1 \sim 10.5$ meV and $\Delta_2 \sim 18$ meV, which are larger than the previous observation in 1-UC FeTe$_{1-x}$Se$_x$[28]. The spatial dependence of the superconducting gap in Fig. 1c measured along an 8 nm line-cut reveals the uniform superconductivity in our ultrahigh crystalline 1-UC FeTe$_{0.5}$Se$_{0.5}$ film. A series of spectra obtained at different temperatures display the gradual rise of the zero-bias conductance (ZBC) from 4.2 K to 45.5 K (Fig. 1d). The SC T$_c$ can be estimated by extrapolating the ZBC in the normalized tunneling spectra to unity[29,30], yielding T$_c$ ~ 62 K as shown in the inset of Fig. 1d.

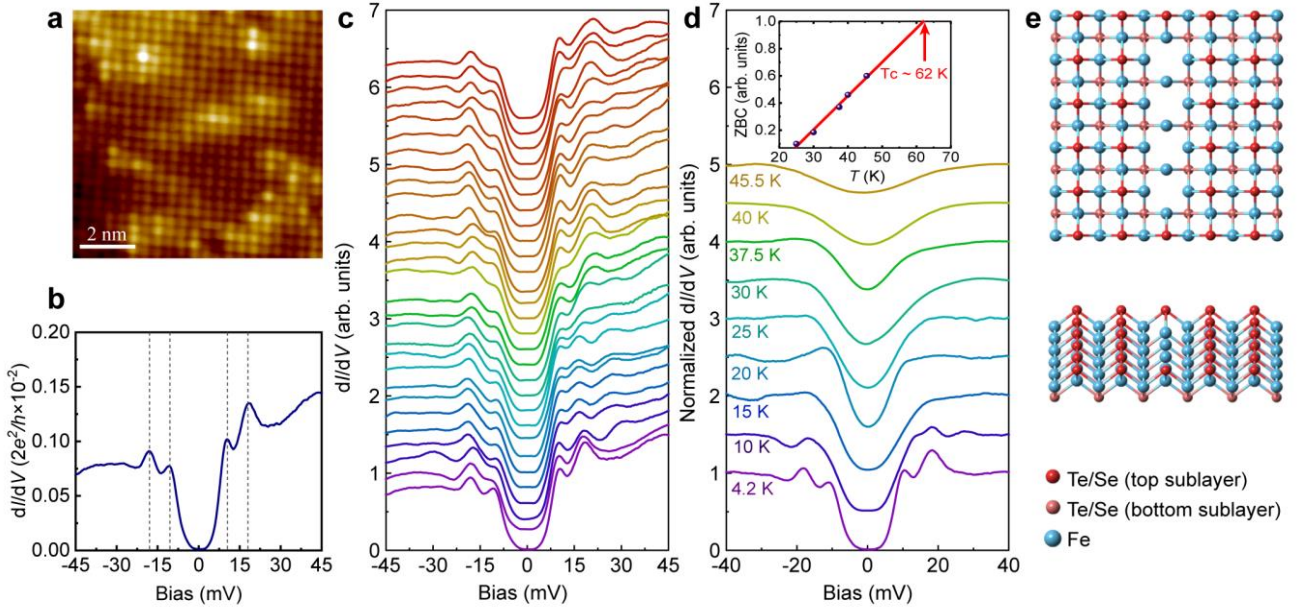

**Figure 1. STM topography and double-gap superconductivity of 1-UC FeTe$_{0.5}$Se$_{0.5}$/STO. a**, STM topographic image of 1-UC FeTe$_{0.5}$Se$_{0.5}$/SrTiO$_3$(001) (9×9 nm$^2$, $V$ = 0.08 V, $I$ = 0.5 nA). **b**, Tunneling spectrum taken on the 1-UC FeTe$_{0.5}$Se$_{0.5}$ film, which shows a U-shaped spectrum with double-gap feature ($V$ = 0.04 V, $I$ = 2.5 nA). **c**, Tunneling spectra taken along an 8 nm line. **d**, Temperature dependence of the normalized tunneling spectra, which are obtained from a cubic fitting to the spectra for $|V| \geqslant 30$ mV. Inset image shows the ZBC extracted from the normalized tunneling spectra, yielding an extrapolated T$_c$ of 62 K at ZBC = 1. **e**, Schematic of the line defects in 1-UC FeTe$_{0.5}$Se$_{0.5}$/STO. The spectra in **c** and **d** are offset vertically for clarify.

In the STM topography of 1-UC FeTe$_{0.5}$Se$_{0.5}$/STO, we can clearly identify 1D atomic line defects. Identifying individual atomic position indicates that the line defect correspond to missing a line of Te/Se atoms as exemplified in Fig. 2a, which is approximately ~ 6 nm long with ~ 15 Te/Se atoms missing. To examine the superconductivity on the atomic line defect, we measure tunneling spectra along the line defects. Remarkably, we observe the ZEBSs emerging at *both* ends of the line defect. Figure 2b shows the spatial map of the tunneling conductance at zero energy around the line defect shown in Fig. 2a, which allows clear visualization of the ZEBSs localized at each end of the defect line. The ZEBS appears to be sharp inside the superconducting gap with the peak amplitude of ~3 times the intensity outside the gap (Fig. 2c). A range of spectra taken along the atomic line defect from the lower end to the upper end shows the spatial distribution of ZEBSs (Fig. 2d). The spectral weight of the ZEBSs decay quickly away from the ends of the chain such that the conductance spectra in the middle of the line defect recover those of the fully gapped superconducting states similar to those obtained far away from the line defect (Fig. 1), albeit with less prominent coherence peaks at the gap edges. These properties are reminiscent of the signatures of Majorana bound states at the ends of the Rashba spin-orbit coupled nanowires[19-22], but with the important difference that the observed ZEBSs are well localized at *both* ends of the line defect and that the measurements were made in zero external magnetic field.

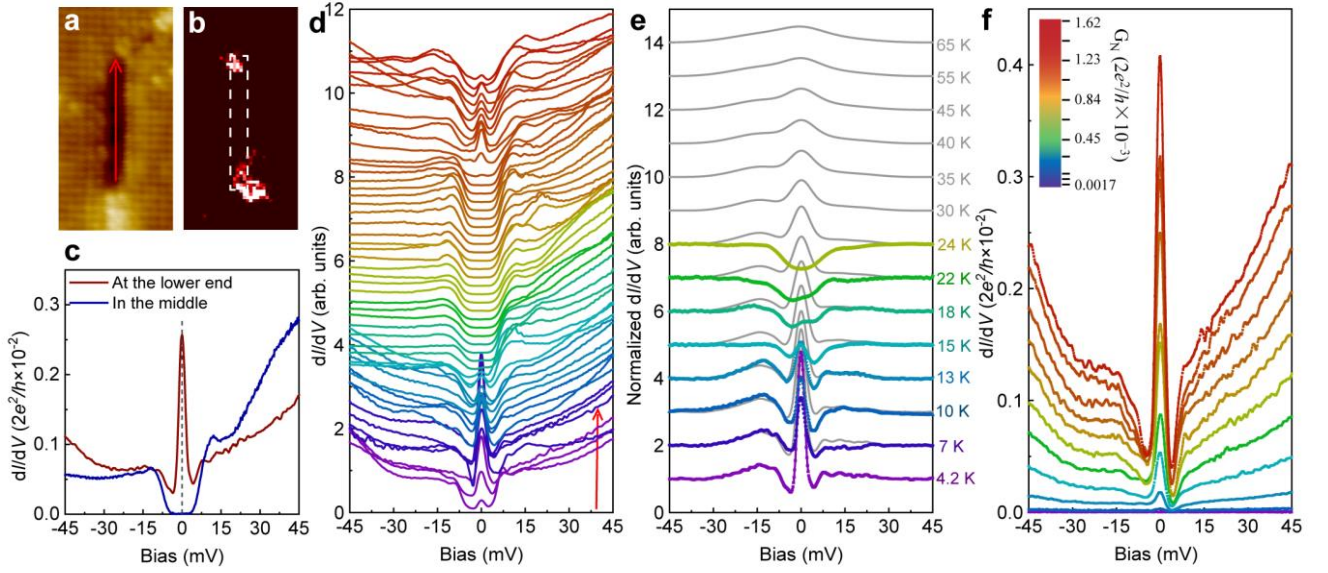

**Figure 2. ZEBS at the ends of a long atomic line defect in 1-UC FeTe$_{0.5}$Se$_{0.5}$/STO. a**, STM topographic image of long atomic line defect on 1-UC FeTe$_{0.5}$Se$_{0.5}$/STO (10.6×5.4 nm$^2$). **b**, Spatial zero-energy mapping at the same area in **a**. The dashed box shows the site of atomic line defect. **c**, Tunneling spectra measured at the lower end and in the middle of the atomic line defect. **d**, Tunneling spectra taken along the red arrow direction in **a**. **e**, Temperature evolution of the ZBCP at the bottom end of the line defect in **a**. The colored curves are normalized tunneling spectra and the grey curves are the convoluted 4.2 K spectra by the Fermi-Dirac distribution function at each temperature. **f**, Tunneling barrier dependence of the ZBCP at the bottom end of the line defect in **a** acquired at 4.2 K. The spectra in **d** and **e** are offset vertically for clarify.

To further investigate the properties of ZEBSs, we perform temperature dependent experiments over the line defects in 1-UC FeTe$_{0.5}$Se$_{0.5}$ film. As the temperature increases, the ZBCP broadens rapidly and reduces in intensity, and finally vanishes completely around 24 K, a temperature far below the transition temperature T$_c$ ~ 62 K (Fig. 2e) where the pairing gap is still quite large. Thus the emergence of the zero-energy end states is a property of the well-developed phase coherent superconducting state at low temperatures. In contrast, ordinary impurity-induced bound states and Kondo resonances would exhibit the Fermi-Dirac thermal broadening behavior and persist to higher temperatures comparable to T$_c$, which is inconsistent with our experimental results. Figure 2f illustrates the evolution of ZEBS under different tunneling barrier conductances, $G_N = I/V$ ($I$ and $V$ are the set points of tunneling spectra), measured at 4.2 K. As the STM tip approaches closer to the sample surface, the ZEBS appears robust and unsplit with the ZBC peak height increases rapidly. More significantly, the ZEBS peak conductance saturates at a relatively high $G_N$ and the quantized conductance can be expected when the experimental temperature is ultralow enough. (Extended Data Fig. 5). The ZBC at different temperatures and under different tunnel barrier conductances follows the scaling behavior under the assumption of tunneling into a single Majorana zero mode[31] (Extended Data Fig. 3-5). Although the lowest experimental temperature is 4.2 K for our STM system, which is relatively high compared to zero temperature, the observed plateau conductance of the ZBC is already on the same order of magnitude as the quantized conductance value[32,33].

Figure 3a shows a shorter (~ 3 nm) atomic line defect, which is about half the length of the one shown in Fig. 2a and has ~ 8 Te/Se atoms missing. The spatial zero-energy map (Fig. 3b) and the conductance spectra at both ends (Fig. 3c) confirm that ZEBSs survive even in this short line defect within our experimental resolution. There are, however, intriguing differences compared to the long line defect shown in Fig. 2. First, the quality factor of the ZEBSs decreases, as clearly seen from the reduced ZBCP height at the end of the line (Fig. 3c). Similarly, the ZBCP disappears entirely at a lower temperature of 19 K (Fig. 3e) compared to 24 K in the case of the longer line defect. Second, ZBCPs with much smaller amplitudes remain even in the middle section of the short line defects (Fig. 3c) inside the somewhat reduced the superconducting gaps, in contrast to the fully gapped superconducting spectrum for the longer line defect shown in Fig. 2c. These features are clearly observable in the tunneling spectra (Fig. 3d) taken along the line defects. They are overall consistent with the presence of coupling between the ZEBSs at the ends that decays exponentially with the length of the chain. The enhanced coupling on shorter line defects are expected to cause the reduction in the zero-energy conductance peak heights and the splitting of the ZEBSs around zero bias. The positive correlation between the ZBC heights and line defect lengths can be deduced from the statistics of the observed ZEBSs in different line defects (Extended Data Fig. 8). We note that within our instrument resolution we did not observe the splitting of the ZBCPs at the lowest measurement temperature of 4.2 K. The fairly robust ZEBS at the ends of the shorter chain can be detected likewise against variations in the tunneling barrier (Fig. 3f).

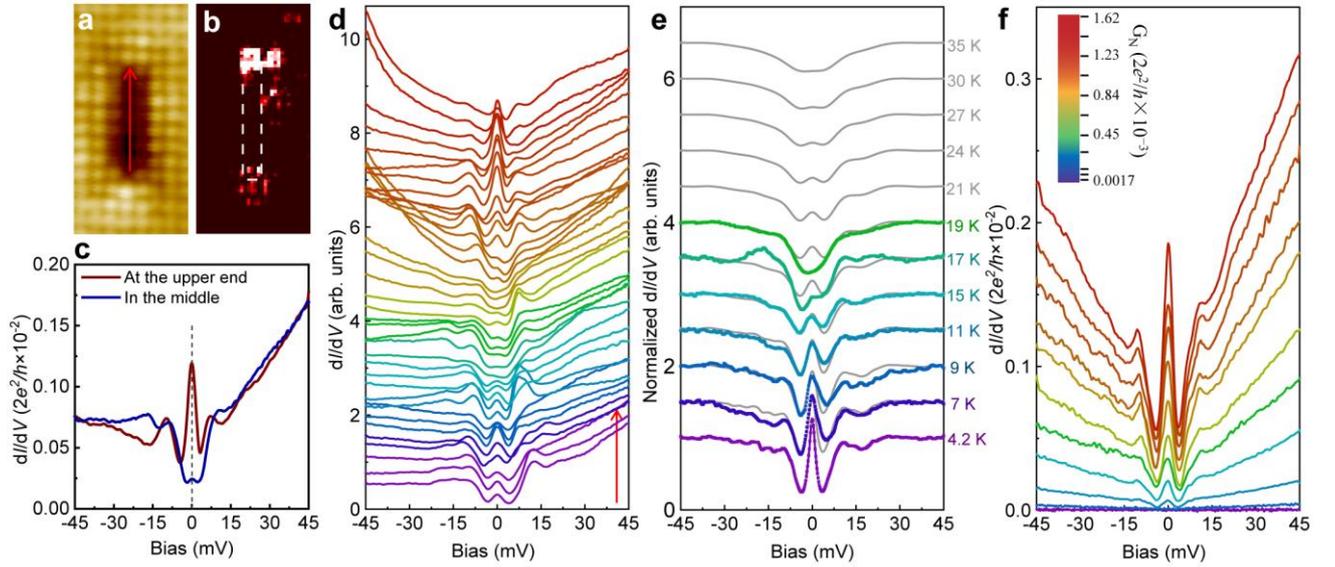

**Figure 3. ZEBS at the ends of a short atomic line defect in 1-UC FeTe$_{0.5}$Se$_{0.5}$/STO. a**, STM topographic image of the shorter atomic line defect on 1-UC FeTe$_{0.5}$Se$_{0.5}$/STO (6.2×3.1 nm$^2$). **b**, Spatial zero-energy mapping at the same area in **a**. The dashed box shows the site of atomic line defect. **c**, Tunneling spectra measured at the upper end and in the middle of the atomic line defect. **d**, Tunneling spectra taken along the red arrow direction in **a**. **e**, Temperature evolution of the ZBCP at the bottom end of the line defect in **a**. The colored curves are normalized tunneling spectra and the grey curves are the convoluted 4.2 K spectra by the Fermi-Dirac distribution function at each temperature. **f**, Tunneling barrier dependence of the ZBCP at the top end of the line defect in **a** acquired at 4.2 K. The spectra in **d** and **e** are offset

vertically for clarify.

The experimental observations suggest the emergence of the ZEBSs as a superconducting analog of the Shockley "surface states" at the ends of the one-dimensional line defects. Indeed, it has been proposed that topologically protected Majorana-Shockley ZEBSs would arise at the ends of an electrostatic line defect in chiral p+ip topological superconductors[27]. In view of the lack of experimental evidence for such topological superconductors in 1-UC FeTe$_{0.5}$Se$_{0.5}$, we outline a different scenario, where the superconducting state is of the non-topological spin-singlet s-wave type, yet the ZEBSs emerge at the ends of the line defect without applying external magnetic field. The band structure of monolayer Fe(Te,Se) has been measured by angle-resolved photoelectron spectroscopy (ARPES) experiments[34] as a function of the heavy Te atom concentration that tends to push down the unoccupied band derived from the Te/Se $p_z$ orbital (hybridized with the Fe $d_{xy}$ orbital) toward the Fermi level. For the current 1-UC FeTe$_{0.5}$Se$_{0.5}$, the $p_z$ band resides just above the chemical potential as schematically shown in Extended Data Figure 9 such that there are no Fermi surfaces near the Γ point at the zone center. The most important effects of the missing Te$^{2-}$/Se$^{2-}$ ions along the line defect are two-fold. First, there is an electrostatic defect potential due to the missing negatively charged Te/Se ions that increases the local electron concentration and pushes the local chemical potential into the $p_z$ band. Second, the missing chain of the Te/Se atoms in the top sublayer of the 1-UC FeTe$_{0.5}$Se$_{0.5}$ causes inversion symmetry breaking around the line defect, which cannot be repaired due to the broken translation symmetry. The resulting electrostatic field perpendicular to the monolayer film leads to an enhanced Rashba spin-orbit interaction that can be significant along the line defect, in addition to the overall contribution due to inversion symmetry breaking by the substrate. We can therefore write down a simple effective model Hamiltonian for the quantum states of such a *Rashba line defect*, which contains a hopping term for the 1D band along the line defect, a RSOC term, and a pairing potential induced by the s-wave superconductivity outside the line defect. The details of the model and analysis are given in Methods. The spin-degenerate 1D Fermi point splits into two by the RSOC. As shown in the Methods, when the zero of the pairing gap function is located between the split Fermi point, the product of the SC gap values at the Fermi level is negative and satisfies the condition for a 1D time-reversal invariant (TRI) topological superconductor[35-38]. Thus, the observed line defect in 1-UC FeTe$_{0.5}$Se$_{0.5}$ may have potentially realized a 1D TRI topological superconductor, which was proposed to arise in hybrid structures of an s-wave superconductor proximity coupled to a Rashba quantum wire[39]. The calculated energy spectrum shows a Kramers pair of Majorana zero modes at each end of the line defect protected by time-reversal symmetry, which manifest as a robust ZBCP in the local density of states at the ends of the chain (Extended Data Figure 9f), in qualitative agreement with our observations. Note that we have ignored the electron bands which contribute to the majority of the density of states and are responsible for the superconducting gaps. However, since the two electron Fermi pockets nearly overlap at the zone corner, resulting in a nearly 4-fold degenerate Fermi point when projected onto the 1D line defects, the condition for the TRI topological superconductor is not affected by the electron bands.

However, we cannot completely rule out the possibility that the time-reversal symmetry turns out to be broken along the line defect directly from our STM data. It is possible that the Fe atoms in the middle sublayer along the defect line become magnetic as the missing Te/Se atoms in the top sublayer remove half of the spin-orbit coupled Fe-chalcogen bonds. A sufficiently strong Zeeman field could thus make the atomic line defect an analog of the hybrid quantum wire[1,2,19-22] or magnetic Fe-chain[40] structures with a single Shockley MZM located at each end. We hope that further experiments, such as spin-polarized STM measurements[41], would clarify this issue definitively. As mentioned above, the explanations of conventional impurity states and Kondo resonances for ZEBSs are inconsistent with our temperature and tunneling barrier dependence measurements. Furthermore, the ZEBSs' dependence on the length of the atomic line defects and their interaction when the ends of two atomic line defects are in close proximity (Extended Data Fig. 6) are difficult to fit into the conventional impurity and Kondo physics, but appear rather naturally in the MZMs interpretation. Additionally, in nodal superconductors, Andreev zero-energy bound states can appear at the surface parallel to the nodal direction, like (110) surface of high-$T_c$ cuprates[42-46]. However, previous ARPES and STM results[28,34] show quite convincingly that the monolayer Fe(Te,Se) is a fully gapped superconductor without nodes. And the ZEBSs we detected here are localized point-like excitations different from a sample "surface". Therefore, we can rule out the possibility of Andreev zero-energy bound states formed at specially directed surfaces in nodal superconductors.

Our observations of ZEBSs at each end of an atomic line defect in 1-UC FeTe$_{0.5}$Se$_{0.5}$/STO provide a novel pathway of searching for Majorana bound states in 2D superconductors. In addition to being a single material with high-$T_c$ and large superconducting gap as well as without applying an external magnetic field, a crucial advantage of our system is the concurrent appearance of ZEBSs at both ends of the line defect, consistent with the inseparable partners of Majorana bound states, which is rare in other 1D quantum structures potentially hosting Majorana end states[47]. This enables us to make a preliminary study of the coupling between the zero-energy end states for different defect chain lengths. Our findings are consistent with and support the interpretation of the observed ZEBSs as Shockley Majorana end states and point to using Rashba line defect as a new platform for studying their statistics and correlations in connection to fault-tolerant topological quantum computing.

**Acknowledgements**

This work was supported by National Natural Science Foundation of China (Grant Nos.11888101，11774008 and 11574134), the National Key Research and Development Program of China (Grant Nos. 2018YFA0305604, 2017YFA0303302 and 2016YFA0300401), the Strategic Priority Research Program of Chinese Academy of Sciences (Grant No. XDB28000000), the Beijing Natural Science Foundation (Grant No. Z180010) and the U.S. Department of Energy, Basic Energy Sciences (Grant No. DE-FG02-99ER45747: K.J. and Z.W.).

**Author Contributions**

J.W. conceived and instructed the research. C.C. grew the samples and analyzed the experimental data. C.C., C.L. and Y.L. carried out the STM/STS experiments. K.J., Y.Z. and Z.W. proposed the theoretical model and performed the theoretical analysis and calculations. C.C., Z.W. and J.W. wrote the manuscript with comments from all authors.

**Additional Information**

Reprints and permissions information is available at www.nature.com/reprints. The authors declare no competing financial interests. Correspondence and requests for materials should be addressed to J.W. (jianwangphysics@pku.edu.cn).

## Methods

**Sample Growth.** Our experiments were performed in an ultrahigh-vacuum (~$2\times10^{-10}$ mbar) MBE-STM combined system (Scienta Omicron). The Nb-doped SrTiO$_3$(001) (wt 0.7 %) substrates were thermally boiled in 90 ℃ deionized water for 50 minutes and then chemically etched in 10% HCl for 45 minutes. Followed by Se-flux method[48] in MBE chamber, the substrates finally obtain the atomically flat TiO$_2$-terminated surface. The 1-UC FeTe$_{0.5}$Se$_{0.5}$ films were grown by co-evaporating high-purity Fe (99.994%) Se (99.999%) and Te (99.999%), with the substrates held at 340 ℃. Then the as-grown 1-UC FeTe$_{0.5}$Se$_{0.5}$ films were annealed at 380 ℃ for 3 hours.

**STM/STS Experiments.** All STM/STS data were acquired in the *in-situ* STM chamber with a polycrystalline PtIr tip by using the standard lock-in technique. In our experiments, the zero-bias offset in tunneling measurements has been calibrated by using the standard method based on the fact that the current should be zero when the voltage is zero. In this work, the zero-energy or zero-bias is defined within the energy resolution of our STM system. The modulation voltage is $V_{mod}$ = 1 mV at 1.7699 kHz, except for that in Extended Data Figure 5 where $V_{mod}$ = 0.5 mV was used. The setup of all STM/STS measurements is $V$ = 0.08 V, $I$ = 0.5 nA for topographic images, and $V$ = 0.04 V, $I$ = 2.5 nA for tunneling spectra unless specified otherwise. All STM/STS measurements were taken at 4.2 K unless specified.

**Discussion on the defects.** In 1-UC FeTe$_{0.5}$Se$_{0.5}$/STO, we observed two different types of Te/Se vacancy defects: 1) Type I shows the atomic dislocations with half-unit-cell shift near the Te/Se vacancy defects. 2) Type II simply shows straight lines of missing Te/Se atoms on the top surface. Statistically, we have about 60% success rate in observing a ZEBS at the end of straight atomic line defects (type II defects with the length less than 10 nm) studied. The success rate may depend on the spatial homogeneity of superconducting gaps and the atomic environment around the line defects.

**Thermal broadening convolution procedure.** We first deconvolute the 4.2 K measured spectrum by using the Fermi-Dirac distribution function to obtain the 4.2 K raw signal without the finite-temperature broadening effect. Then we convolute the 4.2 K raw signal with the Fermi-Dirac distribution function at each high temperature. Finally, we obtain the spectra in responses to the expected thermal broadening at different temperatures.

**Model and Analysis.** Extended Data Figure 9a shows a schematic of the line defect in real space with 3 missing Te/Se atoms in the top sublayer. The band structure of 1-UC FeTe$_{0.5}$Se$_{0.5}$ measured by ARPES[34] near the Γ point is shown in Extended Data Figure 9b. There are two hole-like bands bellow and one electron band of predominately $p_z$ character just above the Fermi level. These will be referred to as the bulk bands. As discussed in the main text, the missing Te/Se atoms along the line defect causes local electron doping and moves the local chemical potential into the electron band (Extended Data Figure 9c). We consider the following effective Hamiltonian for the entire system

$$H = H_t + H_\Delta + H_{soc}$$

where the hopping part is $H_t = \sum_{ij} c_i^+ (t_{ij} - (\epsilon_i + \mu)\delta_{ij})c_j$ with hopping integral $t_{ij}$, onsite energy $\epsilon_i$ and a global chemical potential $\mu$. For simplicity, we set nearest neighbor hopping $t_1 = 0$, and use the second nearest neighbor hopping $t_2 = -1.0$ as the energy unit, and set $\mu = -4.5$. With $\epsilon_i = 0$ away from the line defect, the bulk electron band is shown in Extended Data Figure 9d. The second term in the Hamiltonian H describes spin-singlet paring $H_\Delta = \sum_{ij} \Delta_{ij}(c_{i\uparrow}^+ c_{j\downarrow}^+ - c_{i\downarrow}^+ c_{j\uparrow}^+ + h.c.)$.

For Fe-based superconductors, the latter is dominated by the second nearest neighbor pairing ($s\pm$), such that $\Delta(k) = 4\Delta_0 Cosk_x Cosk_y$. We set $\Delta_0 = 0.1$ globally and note that the specific pairing gap function will not change our results qualitatively. On the line defect, the hopping $t_2$ along the defect line direction is significantly weakened due to the missing Te/Se atoms, which is set to zero for simplicity, leaving only the hopping $t_2$ in the direction perpendicular to the chain. The electrostatic energy is chosen to be $\epsilon_i = 3.9$ such that the defect band crosses the Fermi level (Extended Data Figure 9e). The last term is the Rashba spin-orbit coupling for the Rashba line defect exclusively: $H_{SOC} = \sum_{\ll ij \gg} \lambda(c_{i\uparrow}^+ c_{j\downarrow} - c_{i\downarrow}^+ c_{j\uparrow} + h.c.)$ where the sites i, j runs only along the line defect. We use $\lambda = 0.2$. Due to the Rashba coupling, the spin-degenerate 1D defect band splits into two separated bands. Correspondingly, each Fermi points ($k_F$) splits into $k_+$ and $k_-$ as shown in Extended Data Figure 9e. The condition for a TRI topological superconductor is satisfied when there is one nodal point of $\Delta(k)$ inbetween $k_+$ and $k_-$, such that $\Delta(k_+)\Delta(k_-) < 0$, and a Kramers pair of Majorana zero modes would emerge at each end of the line defect[37,39]. We diagonalize the Hamiltonian in real space on a rectangular sample with its long side ($L_x = 100$ and open boundary) oriented along the line defect direction, short side ($L_y = 20$) wrapped around with the periodic boundary condition, and a defect line ($L_d = 50$) positioned in the middle. The calculated eigenstates energies are shown in the inset in Extended Data Figure 9f, where two Kramers pairs (red and blue dots) of zero-energy Majorana modes are separated from the continuum by the pairing gap are localized at each end of the line defect. The tunneling local density of states at the end of the line defect is shown in Extended Data Figure 9f exhibiting the zero-energy conductance peak due to the Kramers pair of Majorana zero modes protected by the time reversal symmetry.

**Data Availability Statement.**

All data analyzed to evaluate the findings within the paper are available from the authors upon reasonable request.

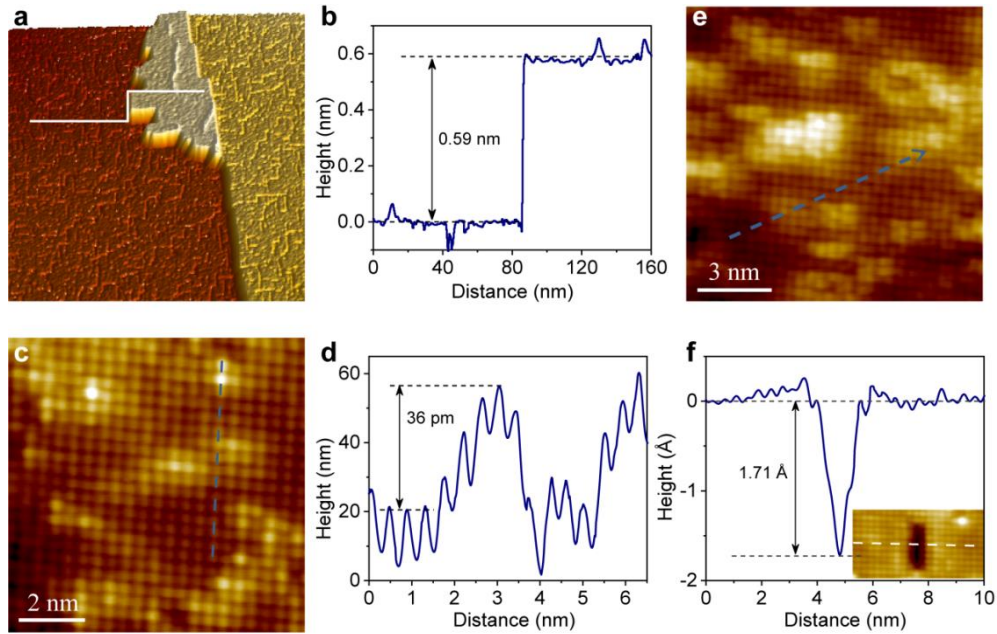

**Extended Data Figure 1. More information about the 1-UC FeTe$_{0.5}$Se$_{0.5}$/STO. a,** The large-scale STM topography of 1-2 UC FeTe$_{0.5}$Se$_{0.5}$/STO with terraces inherited from the STO substrate. The brown and yellow areas are both 1-UC FeTe$_{0.5}$Se$_{0.5}$ film but formed on different STO terraces, and the lighter color means the higher height. The lighter yellow area in the middle shows the 2$^{nd}$-UC FeTe$_{0.5}$Se$_{0.5}$ film grown on the edge of the lower STO terrace, and the white bulge patches are the grain boundaries in the 2$^{nd}$-UC FeTe$_{0.5}$Se$_{0.5}$ film. **b,** Profile taken along the white curve in **a**. The thickness of the 2$^{nd}$-UC FeTe$_{1-x}$Se$_x$ changes with the stoichiometry and its value is about 0.59 nm when x=0.5 (ref 28). **c,** Atomically resolved STM image of 1-UC FeTe$_{0.5}$Se$_{0.5}$/STO (9×9 nm$^2$). **d,** Profile taken along the blue line in **c**. The surface corrugation between the darker and brighter atoms is about 36 pm, which is comparable with the previous results[28]. **e,** The STM image of 1-UC FeTe$_{0.5}$Se$_{0.5}$/STO (12×12 nm$^2$). Spectra in Fig. 1c from the bottom to top are taken along the blue arrow in this image. **f,** Profile taken along the white line across the line defect, as marked in the inset. Considering the anion height away from the Fe-Fe plane (h$_{Te}$ = 1.72 Å and h$_{Se}$ = 1.48 Å in FeSe$_{0.44}$Te$_{0.56}$ crystal[49]), the depth of 1.71 Å in the atomic line defect is identified as the missing of apical chalcogen atoms.

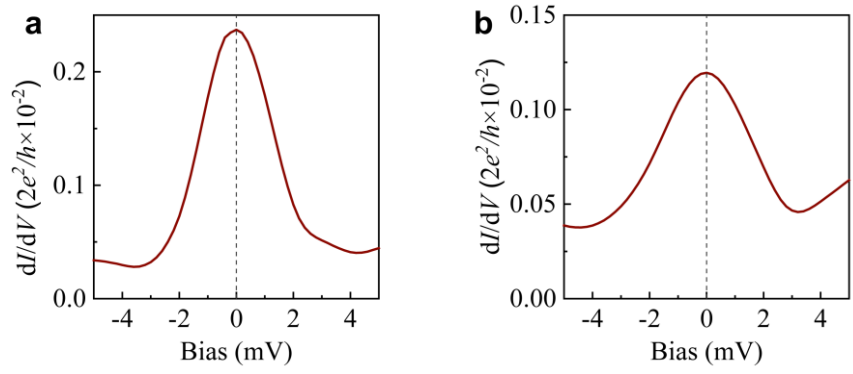

**Extended Data Figure 2. The zoom-in curves of ZEBSs in atomic line defects. a,b,** The zoom-in spectra of ZEBSs in Fig. 2c and Fig. 3c, respectively. The peaks are precisely at zero-bias as the dashed line guiding for eyes.

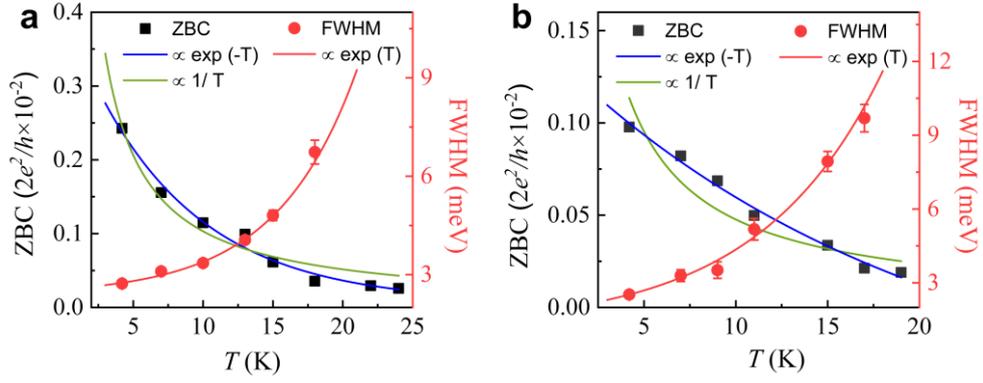

**Extended Data Figure 3. Additional results of temperature evolution. a,b,** Extracted ZBC and FWHM using Lorentzian fitting as a function of temperature from Fig. 2e and Fig. 3e, respectively. With increasing the temperature, the ZBC decreases and the FWHM broadens simultaneously far below $T_c$. The exponential temperature dependence of ZBC is substantially different from the 1/ T behavior typically expected from a normal Andreev bound state, which is evidence against the usual Andreev bound state interpretation and further supports our conclusion[50].

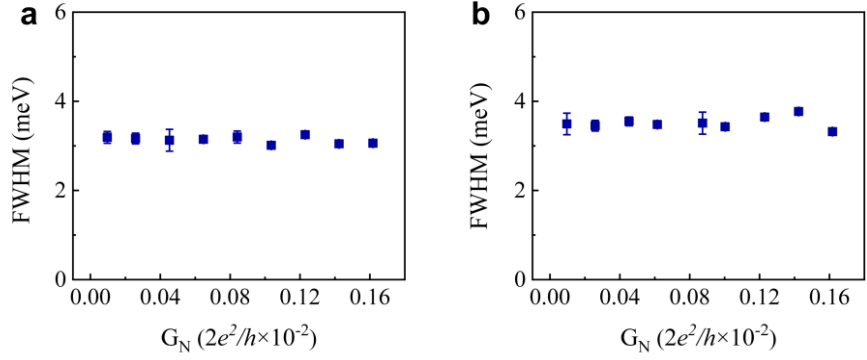

**Extended Data Figure 4. Additional results of tunneling barrier evolution. a,b,** FWHM of ZEBSs at 4.2 K under different tunneling barriers in Fig. 2f and Fig. 3f, respectively. The FWHM and error bars are defined by Lorentzian fitting to the curves. As the STM tip approaches closer to the sample surface, the ZEBS width remains approximately unchanged in the case of weak tunnel coupling. The FWMH of the experimentally detected ZEBSs is almost limited by the instrumental and thermal broadening.

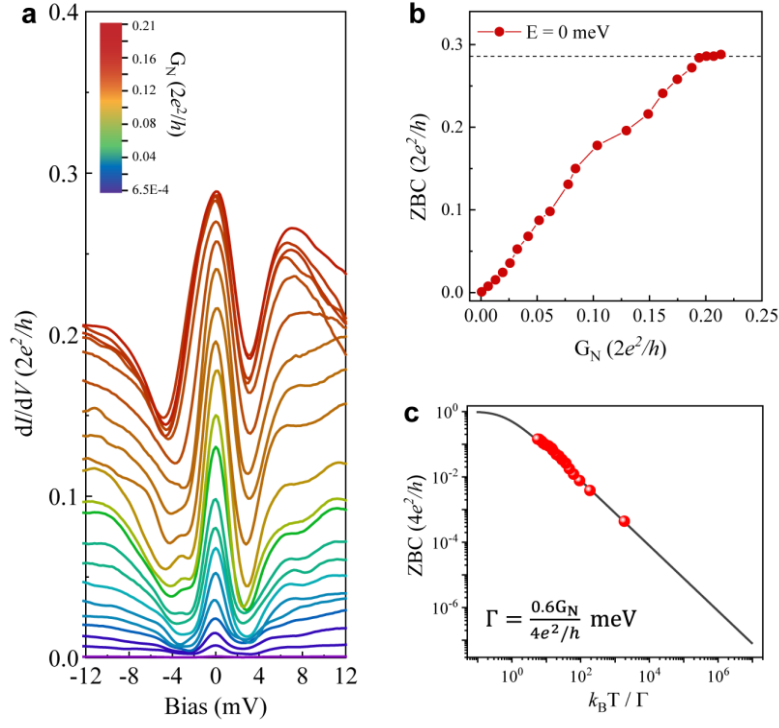

**Extended Data Figure 5. Further tip-approaching-sample measurement. a,** A series of d$I$/d$V$ spectra under different tunneling barriers $G_N$ at 4.2 K. The strong tunnel coupling ($G_N > 0.15 \times 2e^2/h$) broadens the ZEBS width over a high background conductance[22]. **b,** The zero-bias conductance as a function of $G_N$ extracted from **a**, showing the plateau conductance value of $0.285 \times 2e^2/h$. We note that the ZBC plateau conductance at large $G_N$ is on the same order of magnitude but still lower than $2e^2/h$ from a MZM (or $4e^2/h$, if there are two MZMs at one end). This is due to the relatively high experimental temperature (4.2 K), which broadens and lowers the ZBC conductance in the strong tunnel coupling limit. **c,** Scaling behavior of the ZBC shown in **a**. Solid curves: Calculated ZBC of two MZMs with different tunneling barriers at finite temperature according to the scaling function[13,32]. Red balls: ZBC of experimental data at 4.2 K.

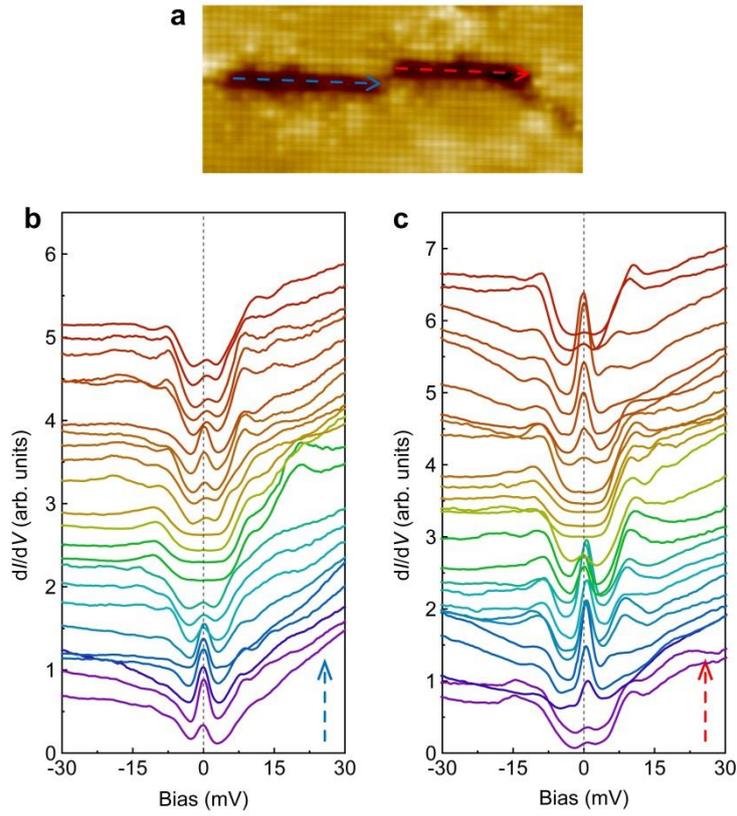

**Extended Data Figure 6. Two neighbouring atomic line defects in 1-UC FeTe$_{0.5}$Se$_{0.5}$ film. a,** Topographic image of two line defects close in space (8×18 nm$^2$). **b,c,** Tunneling spectra taken along the blue and red arrows in **a**, respectively. The energies of the conductance peaks at the two adjacent ends have shifted slightly away from zero due to the coupling between the ZEBSs, while the ZEBSs at the other two far ends are still located at zero.

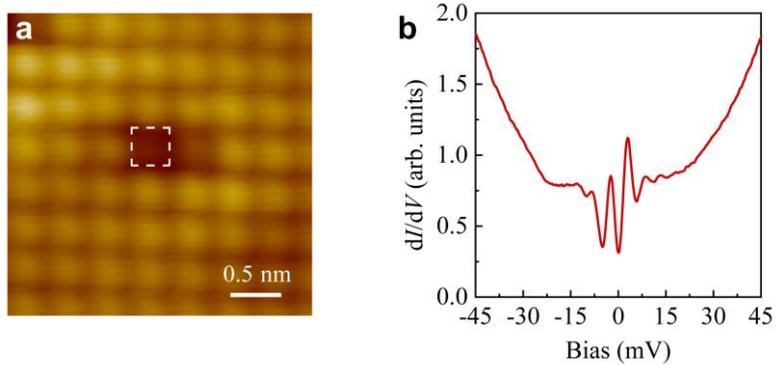

**Extended Data Figure 7. Single Te/Se vacancy defect in 1-UC FeTe$_{0.5}$Se$_{0.5}$/STO. a,** Topographic image of the single Te/Se vacancy defect (3×3 nm$^2$). **b,** Tunneling spectrum measured on the defect site in **a** shows a pair of in-gap bound states, which is different from the ZBCPs at the ends of the atomic line defects.

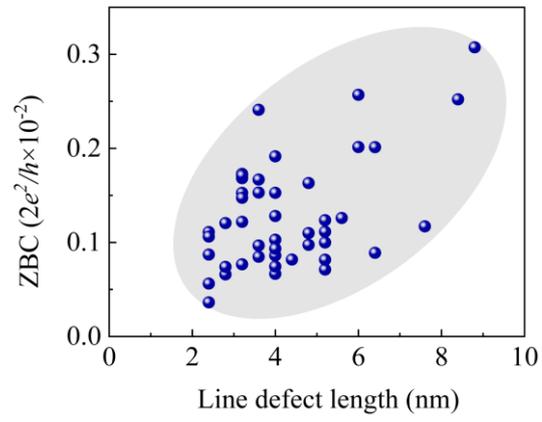

**Extended Data Figure 8. Statistics on the length of line defects vs the heights of the ZEBSs.** We note that there shows a positive relationship between the line defect lengths and the ZBC heights, which suggests the coupling of the two ZEBSs at both ends suppresses the ZBC heights in the case of the shorter line defects.

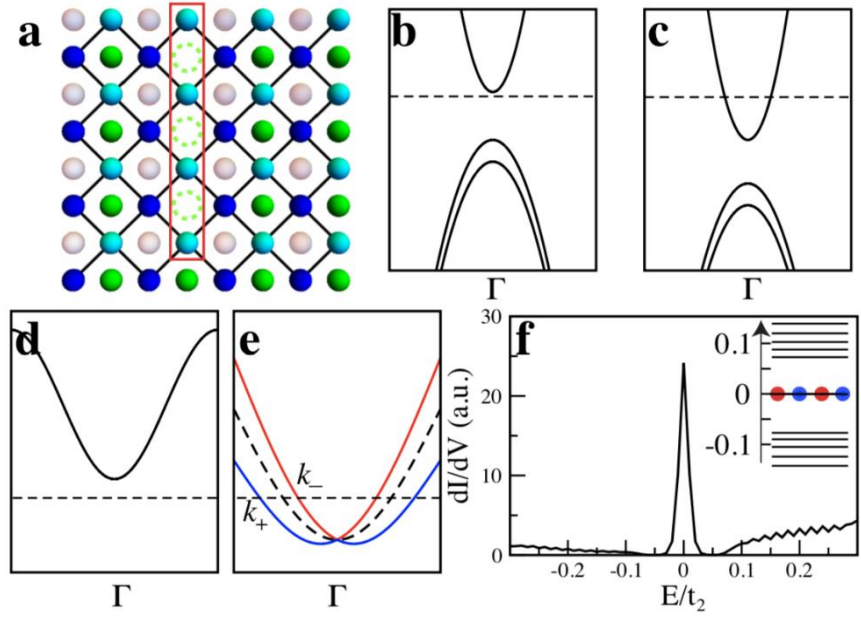

**Extended Data Figure 9. a,** Lattice configuration of FeTe$_{0.5}$Se$_{0.5}$ with a line defect. The blue and cyan balls are two kinds of Fe sites due to different Te/Se crystal fields. The light pink ball is the bottom sublayer of Te/Se atoms while the green balls label the Te/Se atoms in the top sublayer. The line defect is formed by removing a line of Te/Se atoms in the top sublayer, as labeled by dashed green circles. **b,** Band structure observed by ARPES in 1-UC FeTe$_{0.5}$Se$_{0.5}$ around the zone center Γ point[34]. **c,** The electrostatic potential on the line defect induces local electron doping and pushes the electron band down to cross the Fermi level (dashed line). **d,** The bulk electron band away from the line defect in the model plotted with momentum along the direction of the line defect. **e,** Defect band in the model with momentum along the direction of the line defect. Due to RSOC, the Fermi points are split into $k_+$ and $k_-$ pairs. **f,** Calculated tunneling density of states at either end point of the line defect. The sharp conductance peak at zero energy corresponds to the Kramers pair of Majorana zero modes located at the end of the line defect. The inset is the eigenvalue spectrum with four Majorana zero modes forming two Kramers pairs.